\documentclass[aps,prb,showpacs,amsmath,amssymb]{revtex4}
\usepackage{graphicx}
\usepackage{bm}

\def\br{ \bm{r} }
\def\bk{ \bm{k} }
\def\bq{ \bm{q} }

\def\bA{ \bm{A} }
\def\bD{ \bm{D} }
\def\bgam{ \bm{\gamma} }
\def\im{ \mathrm{Im}\, }

\def\tr{ \,\mathrm{tr}\, }

\begin{document}
\title{Nonuniform states in noncentrosymmetric superconductors}

\author{V. P. Mineev$^1$ and K. V. Samokhin$^2$}

\affiliation{$^{1}$ Commissariat \`a l'Energie Atomique,
DSM/DRFMC/SPSMS, 38054 Grenoble, France\\ $^{2}$ Department of
Physics, Brock University, St. Catharines, Ontario L2S 3A1,
Canada}
\date{\today}

\begin{abstract}
In noncentrosymmetric crystals, nonuniform superconducting states
are possible even in the absence of any external magnetic field.
The origin of these states can be traced to the Lifshitz
invariants in the free energy, which are linear in spatial
gradients. We show how various types of the Lifshitz invariants in noncentrosymmetric
superconductors can be derived from microscopic theory.
\end{abstract}


\maketitle

The discovery of superconductivity in CePt$_3$Si (Ref.
\onlinecite{Bauer04}) has renewed interest, both experimental and
theoretical, in the properties of superconductors without
inversion symmetry. One of the most spectacular differences from
the usual, i.e. centrosymmetric, case is the presence of additional terms
in the Ginzburg-Landau free energy, which are linear in spatial
gradients.\cite{Edel96} These terms lead to a ``helical''
superconducting phase, in which the order parameter is nonuniform
in the presence of an external magnetic field which is coupled
only to the spins of electrons.\cite{Agter03,Sam04,KAS05} That
a nonuniform superconducting state can be created by purely
paramagnetic effects was suggested a long time ago by Larkin and
Ovchinnikov\cite{LO64} and Fulde and Ferrell\cite{FF64} (LOFF). In
contrast to the helical state, the LOFF state appears as a result
of the sign change of the second-order gradient term in the free
energy.

According to Refs. \onlinecite{MS94,SZB04}, the crystal symmetry
might sometimes admit the linear in gradients terms -- the Lifshitz
invariants -- in the free energy, leading to nonuniform superconducting states
even in the absence of magnetic field. We do not intend to provide a comprehensive list
of possibilities for all noncentrosymmetric
crystal symmetries. Instead, our purpose is to use two examples to
show how the Lifshitz
invariants at zero field can be obtained microscopically in noncentrosymmetric superconductors
with a Rashba-type spin-orbit (SO) coupling.
In Sec. \ref{sec: inter single}, the case of a spin-triplet order parameter which transforms
according to a three-dimensional irreducible representation of a cubic point group is discussed, while 
in Sec. \ref{sec: inter mix} we consider a mixture of two representations in a tetragonal crystal.
Throughout the paper we use the units in which $\hbar=k_B=1$, and $e$ denotes the
absolute value of the electron charge.

\section{Interband pairing: Single representation}
\label{sec: inter single}

Our starting point is the following Hamiltonian of noninteracting
electrons in a noncentrosymmetric crystal:
\begin{equation}
\label{H_0}
    H_0=\sum\limits_{\bk}\sum_{\alpha\beta=\uparrow,\downarrow}
    [\epsilon_0(\bk)\delta_{\alpha\beta}+\bgam(\bk)\bm{\sigma}_{\alpha\beta}]
    a^\dagger_{\bk\alpha}a_{\bk\beta}
\end{equation}
where $\bm{\sigma}$ are the Pauli matrices, and the sum over $\bk$
is restricted to the first Brillouin zone. The second term in Eq.
(\ref{H_0}), with $\bgam(\bk)=-\bgam(-\bk)$, describes the
Rashba-type (or antisymmetric) SO coupling of electrons with the crystal lattice,
which is specific to noncentrosymmetric systems. In addition, there is a usual (symmetric) 
SO coupling, which is present even in centrosymmetric crystals. If the latter is included then
$\alpha,\beta$ in Eq. (\ref{H_0}) should be interpreted as pseudospin projections.
The Hamiltonian is diagonalized by a unitary transformation
$a_{\bk\alpha}=\sum_{\lambda}u_{\alpha\lambda}(\bk)c_{\bk\lambda}$,
where
\begin{equation}
\label{Rashba_spinors}
    \displaystyle u_{\uparrow\lambda}(\bk)=
    \sqrt{\frac{|\bgam|+\lambda\gamma_z}{2|\bgam|}},\quad
    \displaystyle u_{\downarrow\lambda}(\bk)=\lambda
    \frac{\gamma_x+i\gamma_y}{\sqrt{2|\bgam|(|\bgam|+\lambda\gamma_z)}},
\end{equation}
with the following result:
\begin{equation}
\label{H_0 band}
    H_0=\sum_{\bk}\sum_{\lambda=\pm}\xi_\lambda(\bk)c^\dagger_{\bk\lambda}c_{\bk\lambda}.
\end{equation}
Here the band dispersion functions are
$\xi_\lambda(\bk)=\epsilon_0(\bk)+\lambda|\bgam(\bk)|$. The
normal-state electron Green's functions can be written as
\begin{equation}
\label{G spin}
    \hat{G}(\bk,\omega_n)=
    \sum_{\lambda=\pm}\hat\Pi_\lambda(\bk)G_\lambda(\bk,\omega_n),
\end{equation}
where
\begin{equation}
\label{Pis}
    \hat\Pi_\lambda(\bk)=\frac{1+\lambda\hat{\bm{\gamma}}(\bk)\bm{\sigma}}{2}
\end{equation}
are the band projection operators ($\hat\bgam=\bgam/|\bgam|$), and
\begin{equation}
\label{band GF}
    G_\lambda(\bk,\omega)=\frac{1}{i\omega_n-\xi_\lambda(\bk)}
\end{equation}
are the Green's functions in the band representation.

In this section we show how one can obtain the Lifshitz invariants
for an order parameter corresponding to a single irreducible
representation of the crystal point group. Let us consider a
purely triplet order parameter, which transforms according to an
irreducible representation $\Gamma$ (of dimensionality
$d_\Gamma$):
\begin{equation}
\label{d kr}
    \bm{d}(\bk,\br)=\sum_{a=1}^{d_\Gamma}\eta_a(\br)\bm{\varphi}_a(\bk),
\end{equation}
where $\bm{\varphi}_a(\bk)=-\bm{\varphi}_a(-\bk)$ are the spin-vector
basis functions, see Ref. \onlinecite{Book}. The order parameter
matrix in the spin (or pseudospin) representation has the form
$\Delta_{\alpha\beta}(\bk,\br)=\bm{d}(\bk,\br)\bm{g}_{\alpha\beta}$,
where $\hat{\bm{g}}=i\hat{\bm{\sigma}}\hat\sigma_2$.

Using the standard formalism, see e.g. Ref. \onlinecite{Sam04}, we
obtain for the quadratic terms in the free energy density:
\begin{equation}
\label{F2 gen}
    F_2=\frac{1}{V}\sum_a\eta_a^*\eta_a-\sum_{a,b}\eta_a^*\hat K_{ab}\eta_b,
\end{equation}
where $V>0$ is the coupling constant, and the operator $\hat
K_{ab}$ is obtained from
\begin{equation}
\label{K q}
    K_{ab}(\bq)=\frac{1}{2}T\sum_n\int\frac{d^3\bk}{(2\pi)^3}\varphi^*_{a,i}(\bk)
    \varphi_{b,j}(\bk)\tr\bigl[\hat g_i^\dagger\hat G(\bk+\bq,\omega_n)
    \hat g_j\hat G^T(-\bk,-\omega_n)\bigr]
\end{equation}
by replacing $\bq\to\bD=-i\bm{\nabla}+(2e/c)\bA$. The Matsubara
summation is restricted by the cutoff energy $\omega_c$.
Substituting here expressions (\ref{G spin}) and using the
identity
$$
    \tr\bigl[\hat g_i^\dagger\hat\Pi_{\lambda_1}(\bk+\bq)\hat g_j
    \hat\Pi^T_{\lambda_2}(-\bk)\bigr]=\frac{1-\lambda_1\lambda_2}{2}\delta_{ij}
    +\lambda_1\lambda_2\hat\gamma_i(\bk)\hat\gamma_j(\bk)-
    \frac{i}{2}(\lambda_1-\lambda_2)e_{ijl}\hat\gamma_l(\bk)
$$
(here we neglected the corrections of the order of $q/k_F$), we
obtain:
\begin{eqnarray}
\label{Kab gen}
    K_{ab}(\bq)=\frac{1}{2}T\sum_n\sum_\lambda\int\frac{d^3\bk}{(2\pi)^3}
    \varphi^*_{a,i}(\bk)\varphi_{b,j}(\bk)\bigl\{\hat\gamma_i(\bk)\hat\gamma_j(\bk)
    G_\lambda(\bk+\bq,\omega_n)G_\lambda(-\bk,-\omega_n)\nonumber\\
    +\left[\delta_{ij}-\hat\gamma_i(\bk)\hat\gamma_j(\bk)-i\lambda
    e_{ijl}\hat\gamma_l(\bk)\right]
    G_\lambda(\bk+\bq,\omega_n)G_{-\lambda}(-\bk,-\omega_n)\bigr\}.
\end{eqnarray}
The first (second) line of this expression describes the intraband
(interband) pairing.

The Lifshitz invariants originate from the odd in $\bq$
contribution to the kernel, which in turn comes from the last
(linear in $\hat{\bm{\gamma}}$) term in the second line of Eq.
(\ref{Kab gen}):
\begin{equation}
    K_{ab}^L(\bq)=-\frac{i}{2}q_me_{ijl}T\sum_n\sum_\lambda\lambda
    \int\frac{d^3\bk}{(2\pi)^3}\varphi^*_{a,i}(\bk)
    \varphi_{b,j}(\bk)\hat\gamma_{l}(\bk)v_{\lambda,m}(\bk)
    G^2_\lambda(\bk,\omega_n)G_{-\lambda}(-\bk,-\omega_n).
\end{equation}
Neglecting the difference (of the order of
$|\bm{\gamma}|/\epsilon_F$) between the Fermi velocities in the
two bands, i.e. setting
$\bm{v}_+(\bk)=\bm{v}_-(\bk)=\bm{v}_F(\bk)$, and calculating the
Matsubara sums, we obtain in the vicinity of the critical
temperature:
\begin{equation}
\label{K L gen}
    K_{ab}^L(\bq)=-i\frac{N_F}{4\pi T_{c0}}q_me_{ijl}
    \left\langle\Phi(\bk)
    \varphi^*_{a,i}(\bk)\varphi_{b,j}(\bk)\hat\gamma_{l}(\bk)
    v_{F,m}(\bk)\right\rangle_{\hat{\bk}},
\end{equation}
where the angular brackets denote the Fermi-surface averaging,
$$
    \Phi(\bk)=\im\Psi'\left(\frac{1}{2}+i\frac{|\bm{\gamma}(\bk)|}{2\pi
    T_{c0}}\right)\simeq -\frac{7\zeta(3)|\bm{\gamma}(\bk)|}{\pi
    T_{c0}},
$$
and $\Psi(x)$ is the digamma function. We assume that the Rashba
SO coupling is sufficiently weak in order for the interband
pairing to survive.
We note that in the absence of time-reversal symmetry breaking in
the normal state, the basis functions can be chosen real. Then it
follows from Eq. (\ref{K L gen}) that the Lifshitz invariants are
absent for order parameters transforming according to
one-dimensional representations of the point group.

Let us consider as an example a three-dimensional order parameter
$\bm{\eta}=(\eta_1,\eta_2,\eta_3)$ in a cubic superconductor,
which corresponds to the representation $F_1$ of the point group
$\mathbb{G}=\mathbf{O}$. We assume a spherical Fermi surface and
describe the SO coupling by
\begin{equation}
\label{gamma O}
    \bgam(\bk)=\gamma_0\bk,
\end{equation}
where $\gamma_0$ is a constant. For the band dispersions we have
$\xi_\lambda(\bk)=\epsilon_0(\bk)+\lambda\alpha$, where
$\alpha=|\gamma_0|k_F$ is the measure of the SO band splitting ($k_F$
is the Fermi wave vector). The normalized spin-vector basis
functions have the following form:
\begin{equation}
\label{phis cubic}
    \varphi_{a,i}(\bk)=\sqrt{\frac{3}{2}}e_{aij}\hat k_j.
\end{equation}
Inserting this into Eq. (\ref{K L gen}), we obtain:
$K_{ab}^L(\bq)=i(7\zeta(3)/8\pi^2)(N_Fv_F\alpha/T_{c0}^2)e_{abi}q_i$.
From this it follows that the Lifshitz invariant in the free energy density has the following form:
\begin{equation}
\label{FL cubic}
    F_L=i\tilde K\left(\eta_1^*D_y\eta_3+\eta_2^*D_z\eta_1+\eta_3^*D_x\eta_2-c.c.\right),
\end{equation}
where
\begin{equation}
\label{tilde K cubic}
    \tilde K=\frac{7\zeta(3)N_Fv_F^2}{8\pi^2 T_{c0}^2}\frac{\alpha}{v_F}.
\end{equation}

Note that, according to Eq. (\ref{phis cubic}), the order parameter (\ref{d kr}) satisfies
$\bm{d}(\bk,\br)\perp\bgam(\bk)$. In the band representation, this corresponds to
interband pairing, as opposed to the limit of strong 
SO band splitting, in which only the component
$\bm{d}\parallel\bgam$ survives (intraband or ``protected'' component), in addition to the spin-singlet
component.\cite{FAKS04}

\section{Interband pairing: Mixture of two representations}
\label{sec: inter mix}

In this section we discuss a different mechanism of producing the Lifshitz invariants. 
We start with the pairing interaction in the band representation, which can be written
a general form as follows:
\begin{equation}
\label{H int band gen}
    H_{int}=\frac{1}{2{\cal V}}\sum_{\bk\bk'\bq}\sum_{\lambda_{1,2,3,4}}
    t_{\lambda_2}(\bk)t^*_{\lambda_3}(\bk')
    \tilde V_{\lambda_1\lambda_2\lambda_3\lambda_4}(\bk,\bk')c^\dagger_{\bk+\bq,\lambda_1}
    c^\dagger_{-\bk,\lambda_2}c_{-\bk',\lambda_3}
    c_{\bk'+\bq,\lambda_4},
\end{equation}
where
\begin{eqnarray}
\label{t lambda}
    t_\lambda(\bk)=\lambda
    \frac{\gamma_x(\bk)-i\gamma_y(\bk)}{\sqrt{\gamma_x^2(\bk)+\gamma_y^2(\bk)}}
\end{eqnarray}
are phase factors,
\begin{eqnarray}
\label{tilde V gen}
    \tilde V_{\lambda_1\lambda_2\lambda_3\lambda_4}(\bk,\bk')=
    v_g(\bk,\bk')\delta_{\lambda_1\lambda_2}\delta_{\lambda_3\lambda_4}
    +v_{u,ij}(\bk,\bk')\tau_{i,\lambda_1\lambda_2}(\bk)
    \tau_{j,\lambda_3\lambda_4}(\bk')\nonumber\\
    +v_{m,i}(\bk,\bk')\tau_{i,\lambda_1\lambda_2}(\bk)\delta_{\lambda_3\lambda_4}
    +v_{m,i}(\bk',\bk)\delta_{\lambda_1\lambda_2}\tau_{i,\lambda_3\lambda_4}(\bk'),
\end{eqnarray}
and $\hat\tau_i(\bk)=\hat u^\dagger(\bk)\hat\sigma_i\hat u(\bk)$,
with the matrices $\hat u(\bk)$ defined by Eq.
(\ref{Rashba_spinors}). The functions $v_g$, $v_{u,ij}$, and
$v_{m,i}$ describe the pairing strength and anisotropy in
spin-singlet, spin-triplet, and mixed channels, respectively. We
follow the notations of Ref. \onlinecite{SM08} and assume that the
frequency dependence of the pairing amplitudes is factorized. The
terms with $\lambda_1=\lambda_2$ and $\lambda_3=\lambda_4$
describe intraband pairing and the scattering of the Cooper pairs
from one band to the other, while the remaining terms describe
pairing of electrons from different bands.

Treating the interaction (\ref{H int band gen}) in the mean-field
approximation, one introduces the gap functions
$\tilde\Delta_{\lambda_1\lambda_2}(\bk,\bq)=t^*_{\lambda_2}(\bk)
\Delta_{\lambda_1\lambda_2}(\bk,\bq)$,
with the following symmetry properties:
\begin{equation}
\label{Delta symm}
    \tilde\Delta_{\lambda_1\lambda_2}(\bk,\bq)=\lambda_1\lambda_2
    \tilde\Delta_{\lambda_2\lambda_1}(-\bk,\bq).
\end{equation}
These are related to the symmetry of the pairing interaction:
$\tilde
V_{\lambda_2\lambda_1\lambda_3\lambda_4}(-\bk,\bk')=\lambda_1\lambda_2
\tilde V_{\lambda_1\lambda_2\lambda_3\lambda_4}(\bk,\bk')$, which
can be easily established from Eq. (\ref{H int band gen}).
Near the critical temperature, the gap functions satisfy the
linearized gap equations:
\begin{equation}
\label{gap eq gen}
    \tilde\Delta_{\lambda_1\lambda_2}(\bk,\bq)=-T\sum_n\int\frac{d^3\bk'}{(2\pi)^3}
    \sum_{\lambda_3\lambda_4}\tilde V_{\lambda_1\lambda_2\lambda_4\lambda_3}(\bk,\bk')
    G_{\lambda_3}\left(\bk'+\bq,\omega_n\right)
    G_{\lambda_4}\left(-\bk',-\omega_n\right)
    \tilde\Delta_{\lambda_3\lambda_4}(\bk',\bq),
\end{equation}
where $G_\lambda(\bk,\omega_n)$ are the Green's functions of band electrons, see Eq. (\ref{band GF}).

We describe pairing anisotropy by the following model, which is
compatible with all symmetry requirements:
\begin{eqnarray}
\label{v model}
    &&v_g(\bk,\bk')=-V_g,\nonumber\\
    &&v_{u,ij}(\bk,\bk')=-V_u(\hat\bgam(\bk)\hat\bgam(\bk'))\delta_{ij},\quad\\
    &&v_{m,i}(\bk,\bk')=0,\nonumber
\end{eqnarray}
where $V_g$ and $V_u$ are constants. Then, from Eq. (\ref{tilde V gen}) one
obtains:
\begin{equation}
\label{tilde V model}
    \tilde V_{\lambda_1\lambda_2\lambda_3\lambda_4}(\bk,\bk')=
    -V_g\delta_{\lambda_1\lambda_2}\delta_{\lambda_3\lambda_4}
    -V_u(\hat\bgam(\bk)\hat\bgam(\bk'))(\bm{\tau}_{\lambda_1\lambda_2}(\bk)
    \bm{\tau}_{\lambda_3\lambda_4}(\bk')).
\end{equation}

Further steps essentially depend on the crystal symmetry, which
determines the momentum dependence of the SO coupling. Let us
consider a tetragonal superconductor with the point group
$\mathbb{G}=\mathbf{C}_{4v}$, in which case one can write
\begin{equation}
\label{gamma C4v}
    \bgam(\bk)=\gamma_0(\hat z\times\bk),
\end{equation}
where $\gamma_0$ is a constant. We assume a cylindrical Fermi
surface along the $z$-axis. For the band dispersions we then have
$\xi_\lambda(\bk)=\epsilon_0(\bk)+\lambda\alpha$, where
$\alpha=|\gamma_0|k_F$. From
Eq. (\ref{tilde V model}) we obtain the pairing interaction
components as follows:
\begin{eqnarray}
    &&\tilde V_{++++}=\tilde V_{----}=-V_g-V_uM^2,\quad
    \tilde V_{++--}=\tilde V_{--++}=-V_g+V_uM^2,\nonumber\\
    &&\tilde V_{+-+-}=\tilde V_{-+-+}=-V_uM+V_uM^2,\quad
    \tilde V_{+--+}=\tilde V_{-++-}=-V_uM-V_uM^2,\nonumber\\
    &&\tilde V_{+++-}=\tilde V_{+---}=\tilde V_{-+++}=\tilde
    V_{---+}=iV_uMN,\\
    &&\tilde V_{++-+}=\tilde V_{+-++}=\tilde V_{-+--}=\tilde
    V_{--+-}=-iV_uMN,\nonumber
\end{eqnarray}
where $M=\hat\bk\hat\bk'$ and $N=(\hat\bk\times\hat\bk')_z$. Note
that the interband components in the second line here contain
terms that are even in both $\bk$ and $\bk'$, as well as ones that
are odd in $\bk$ and $\bk'$. As we shall see, this gives rise to
inhomogeneous superconducting states.

The pairing interaction can be presented as an expansion over the
irreducible representations $\Gamma$ of the point group
$\mathbb{G}$. The tetragonal group $\mathbf{C}_{4v}$ has four
one-dimensional representations: $A_1$, $A_2$, $B_1$, $B_2$, and
one two-dimensional representation $E$ (the notations are the same
as in Ref. \onlinecite{LL3}). The simplest polynomial expressions
for the normalized basis functions have the following form:
\begin{equation}
    \varphi_{A_1}(\bk)=\hat k_x^2+\hat k_y^2=1,\quad
    \varphi_{B_1}(\bk)=\sqrt{2}(\hat k_x^2-\hat k_y^2),\quad
    \varphi_{B_2}(\bk)=2\sqrt{2}\hat k_x\hat k_y,\quad
    \bm{\varphi}_{E}(\bk)=\sqrt{2}(\hat k_x,\hat k_y).
\end{equation}
Using
\begin{eqnarray*}
    &&M=\frac{1}{2}\bm{\varphi}_E(\bk)\bm{\varphi}_E(\bk'),\\
    &&M^2=\frac{1}{2}\varphi_{A_1}(\bk)\varphi_{A_1}(\bk')+\frac{1}{4}\varphi_{B_1}(\bk)\varphi_{B_1}(\bk')
    +\frac{1}{4}\varphi_{B_2}(\bk)\varphi_{B_2}(\bk'),\\
    &&MN=\frac{1}{4}\varphi_{B_1}(\bk)\varphi_{B_2}(\bk')-\frac{1}{4}\varphi_{B_2}(\bk)\varphi_{B_1}(\bk'),
\end{eqnarray*}
we obtain from the gap equations (\ref{gap eq gen}) the following expressions for
the intraband gap functions:
\begin{equation}
    \tilde\Delta_{\lambda\lambda}=\eta^{A_1}_{\lambda\lambda}\varphi_{A_1}(\bk)
    +\eta^{B_1}_{\lambda\lambda}\varphi_{B_1}(\bk)+\eta^{B_2}_{\lambda\lambda}\varphi_{B_2}(\bk),
\end{equation}
and also for the interband gap functions:
\begin{equation}
    \tilde\Delta_{+-}=\eta^{A_1}_{+-}\varphi_{A_1}(\bk)+\eta^{B_1}_{+-}\varphi_{B_1}(\bk)
    +\eta^{B_2}_{+-}\varphi_{B_2}(\bk)+\bm{\eta}^{E}_{+-}\bm{\varphi}_E(\bk).
\end{equation}
$\tilde\Delta_{-+}$ can be obtained from $\tilde\Delta_{+-}$
using Eq. (\ref{Delta symm}). The expansion coefficients
$\eta^\Gamma_{\lambda\lambda'}(\bq)$ play the role of the order
parameter components.

In a general nonuniform case and at arbitrary values of the SO
band splitting and the coupling constants $V_g$ and $V_u$ [Eqs. (\ref{v model})], the gap equations (\ref{gap eq gen}) yield a
set of coupled equations for the eleven components of the order
parameter. In order to demonstrate the possibility of
inhomogeneous solutions even in the absence of an external
magnetic field, it is sufficient to retain only the interband
components in the $A_1$ and $E$ channels. To turn off the
intraband pairing, we assume that $V_g<0$ (i.e. the isotropic
channel is repulsive), and introduce the notations
$\eta^{A_1}_{+-}=\eta$ and $\bm{\eta}^{E}_{+-}=\bm{\xi}$. Then,
using the symmetry properties (\ref{Delta symm}), one can write
\begin{equation}
\label{OP model}
    \tilde\Delta_{+-}(\bk,\bq)=\eta(\bq)\varphi_{A_1}(\bk)+\bm{\xi}(\bq)\bm{\varphi}_E(\bk),\quad
    \tilde\Delta_{-+}(\bk,\bq)=-\eta(\bq)\varphi_{A_1}(\bk)+\bm{\xi}(\bq)\bm{\varphi}_E(\bk).
\end{equation}
The linearized gap equations take the form
\begin{equation}
\label{eta xi eqs gen}
    \eta=S^{\eta\eta}\eta+S^{\eta\xi}_a\xi_a,\quad
    \xi_a=S^{\eta\xi}_a\eta+S^{\xi\xi}_{ab}\xi_b,
\end{equation}
where $a,b=1,2$, and
\begin{eqnarray*}
    &&S^{\eta\eta}(\bq)=V_uT\sum_n\int\frac{d^3\bk}{(2\pi)^3}\varphi_{A_1}^2(\bk)G_+(\bk+\bq,\omega_n)
    G_-(-\bk,-\omega_n),\\
    &&S^{\eta\xi}_a(\bq)=V_uT\sum_n\int\frac{d^3\bk}{(2\pi)^3}\varphi_{A_1}(\bk)\varphi_{E,a}(\bk)
    G_+(\bk+\bq,\omega_n)G_-(-\bk,-\omega_n),\\
    &&S^{\xi\xi}_{ab}(\bq)=V_uT\sum_n\int\frac{d^3\bk}{(2\pi)^3}\varphi_{E,a}(\bk)\varphi_{E,b}(\bk)
    G_+(\bk+\bq,\omega_n)G_-(-\bk,-\omega_n).
\end{eqnarray*}

At $\bq=0$, one obtains $S^{\eta\xi}_a=0$, which means that the $A_1$ and $E$ channels
are decoupled. The critical
temperature of the phase transition into a uniform superconducting
state is the same for $\eta$ and $\bm{\xi}$:
\begin{equation}
\label{Tc uniform}
    T_{c0}=T^{BCS}_{c0}-\frac{7\zeta(3)}{16\pi^2T_{c0}}\alpha^2,
\end{equation}
where $T^{BCS}_{c0}=(2e^{\mathbb{C}}/\pi)\omega_c\exp(-1/N_FV_u)$
is the standard BCS expression for the critical temperature in the
absence of SO band splitting (i.e. at $\alpha=0$), $\zeta(x)$ is
the Riemann zeta-function, $\mathbb{C}\simeq 0.577$ is Euler's
constant, and $N_F$ is the density of states at the Fermi level
(we neglect the difference between the densities of states in the
two bands). It is assumed that $\alpha\lesssim T_{c0}$, otherwise
the interband pairing is suppressed by the same mechanism that
suppresses the singlet pairing in centrosymmetric superconductors.
Indeed, the SO band splitting enters the Green's
functions and the gap equations in exactly the same way as the Zeeman
field does in the centrosymmetric case.

The actual critical temperature of the superconducting transition is higher
than $T_{c0}$. Keeping the lowest terms in the gradient expansion
in Eqs. (\ref{eta xi eqs gen}), we obtain:
\begin{equation}
\label{eta xi eqs}
    \begin{array}{l}
    \displaystyle \left[a(T-T_{c0})+Kq^2\right]\eta+\tilde K q_i\xi_i=0, \\
    \displaystyle
    \tilde K q_i\eta+\left[a(T-T_{c0})\delta_{ij}
    +\frac{K}{2}(\delta_{ij}q^2+2q_iq_j)\right]\xi_j=0, \\
    \end{array}
\end{equation}
where $a=N_F/T_{c0}$,
$K=7\zeta(3)N_Fv_F^2/32\pi^2T_{c0}^2$, and $\tilde
K=2\sqrt{2}K\alpha/v_F$. The linear in $\bq$ terms, which mix the
two channels, correspond to the Lifshitz invariant in the
Ginzburg-Landau free energy. In the coordinate representation,
this invariant has the following form:
\begin{equation}
\label{Lif-inv}
    F_L=\tilde
    K\left[\eta^*(\bm{D}\bm{\xi})+(\bm{\xi}^*\bm{D})\eta\right],
\end{equation}
where $\bm{D}=-i\bm{\nabla}$, or, in the presence of magnetic
field, $\bm{D}=-i\bm{\nabla}+(2e/c)\bm{A}$. It is easy to see that
this expression satisfies all symmetry requirements. In
particular, it is invariant under time reversal operation, which
in the band representation is expressed as
$\tilde\Delta_{\lambda_1\lambda_2}(\bk)\to\tilde\Delta_{\lambda_2\lambda_1}^*(\bk)$.
According to Eq. (\ref{OP model}), the order parameter components
transform under time reversal as follows: $\eta\to-\eta^*$,
$\bm{\xi}\to\bm{\xi}^*$.

Seeking the order parameter in the form
\begin{equation}
\label{OP components}
    \eta(\br)=\eta_0 e^{iqx},\quad \bm{\xi}(\br)=\xi_0 e^{iqx}(1,0),
\end{equation}
we obtain the following expression for the critical temperature as
a function of $q$:
\begin{equation}
    a(T-T_{c0})=-\frac{5}{4}Kq^2+\frac{1}{2}Kq\sqrt{32\left(\frac{\alpha}{v_F}\right)^2
    +\frac{1}{4}q^2}
\end{equation}
This function has a maximum at finite $q=q_0$, where
\begin{equation}
\label{q_0}
    q_0=c_1\frac{\alpha}{v_F},
\end{equation}
with $c_1\simeq 1.15$. The corresponding critical temperature is
\begin{equation}
\label{Tc nonuniform}
    T_c=T_{c0}+c_2\frac{K}{a}\left(\frac{\alpha}{v_F}\right)^2,
\end{equation}
where $c_2\simeq 1.62$. Thus, the nonuniform superconducting phase
described by Eq. (\ref{OP components}) has a higher critical
temperature than the uniform state.

It is instructive to interpret our results using the spin
representation of the order parameter:
\begin{equation}
\label{Delta spin gen}
    \Delta_{\alpha\beta}(\bk,\bq)=\sum_{\lambda_1\lambda_2}
    u_{\alpha\lambda_1}(\bk)\Delta_{\lambda_1\lambda_2}(\bk,\bq)u_{\beta\lambda_2}(-\bk)=
    -[\hat u(\bk)\hat{\tilde\Delta}(\bk,\bq)\hat
    u^\dagger(\bk)(i\hat\sigma_2)]_{\alpha\beta}.
\end{equation}
The interband elements, see Eq. (\ref{OP model}), are translated into the spin representation as follows:
\begin{equation}
\label{Delta spin}
    \Delta_{\alpha\beta}(\bk,\bq)=(i\bm{\sigma}\sigma_2)_{\alpha\beta}\bm{d}(\bk,\bq),
\end{equation}
where
\begin{equation}
    d_x=-i\eta(\bq)\varphi_{A_1}(\bk)\hat k_x,\quad d_y
    =-i\eta(\bq)\varphi_{A_1}(\bk)\hat k_y,
    \quad d_z=-\bm{\xi}(\bq)\bm{\varphi}_E(\bk).
\end{equation}
Therefore, the pairing symmetry in our model is purely
spin-triplet, with $\bm{d}(\bk,\bq)\perp\bgam(\bk)$.

The Cooper pairs in the nonuniform state
are composed of electrons from different SO-split bands, and the
order parameter is modulated with the wave vector
$q_0\sim(\alpha/T_{c0})\xi_0^{-1}$ (here $\xi_0$ is the coherence
length). This effect formally resembles the nonuniform
mixed-parity state (NMP) in centrosymmetric superconductors and
superfluids in the presence of magnetic field, which was discussed
in Refs. \onlinecite{Lebed06,NMP}. The reason is that, as was
mentioned above, the SO band splitting in the noncentrosymmetric
case affects the interband pairing of electrons of opposite
helicity in the same way as the Zeeman field in centrosymmetric
superconductors affects the usual BCS pairing between electrons of
opposite spin. However, the NMP state originates from the triplet
interaction channel introduced in the model along with the
singlet channel. In contrast, the inhomogeneous superconducting
state in the noncentrosymmetric case arises from a purely triplet
pairing interaction, see the second line in Eq. (\ref{v model}),
which acquires both $\bk$-even and $\bk$-odd components in the
band representation.

\section{Conclusions}
\label{sec: conclusions}

We come to the conclusion that nonuniform superconducting states
can exist in noncentrosymmetric superconductors even in the
absence of external magnetic field. Experimentally, the zero-field nonuniform states 
discussed in this article can only be
observed in the noncentrosymmetric compounds with the SO band
splitting smaller than the superconducting critical temperature.
To the best of the authors' knowledge, in all noncentrosymmetric
compounds discovered to date the relation between the two energy
scales is exactly the opposite: the SO band splitting exceeds all
superconducting energy scales by orders of magnitude, completely
suppressing the interband pairing, both uniform and nonuniform.

We would like to note that the models considered in Secs.
\ref{sec: inter single} and \ref{sec: inter mix}  are different
from those discussed previously in the literature. For instance, if the order parameter transforms
according to a two-dimensional representation of the point group
$\mathbf{D}_6$ (Ref. \onlinecite{MS94}), or has two intraband
components, $\eta_+$ and $\eta_-$, of the same symmetry in a tetragonal crystal (Ref.
\onlinecite{SZB04}), then the Lifshitz invariant has the following form:
\begin{equation}
\label{Lif-inv-old}
    F_L=i\tilde K(\eta_+^*D_z\eta_--\eta_-^*D_z\eta_+),
\end{equation}
where $\tilde K$ is a real constant. Although such terms satisfy
all symmetry requirements and are therefore possible on
phenomenological grounds, they are absent in our microscopic
models. Moreover, expressions of the form (\ref{Lif-inv-old}) should exist
even for the large SO band splitting, in which case the interband
pairing channels are suppressed, and the mechanisms discussed in Secs.
\ref{sec: inter single} and \ref{sec: inter mix} do not work.


\begin{thebibliography}{99}

\bibitem{Bauer04}
E. Bauer, G. Hilscher, H. Michor, Ch. Paul, E. W. Scheidt, A.
Gribanov, Yu. Seropegin, H. No\"el, M. Sigrist, and P. Rogl, Phys.
Rev. Lett. \textbf{92}, 027003 (2004).

\bibitem{Edel96}
V. M. Edelstein, J. Phys.: Condens. Matter \textbf{8}, 339 (1996).

\bibitem{Agter03}
D. F. Agterberg, Physica C \textbf{387}, 13 (2003).

\bibitem{Sam04}
K. V. Samokhin, Phys. Rev. B \textbf{70}, 104521 (2004).

\bibitem{KAS05}
R. P. Kaur, D. F. Agterberg, and M. Sigrist, Phys. Rev. Lett.
\textbf{94}, 137002 (2005).

\bibitem{LO64}
A. I. Larkin and Yu. N. Ovchinnikov, Zh. Eksp. Teor. Fiz.
\textbf{47}, 1136 (1964) [Sov. Phys. JETP \textbf{20}, 762
(1965)].

\bibitem{FF64}
P. Fulde and R. A. Ferrell, Phys. Rev. \textbf{135}, A550 (1964).

\bibitem{MS94}
V. P. Mineev and K. V. Samokhin, Zh. Eksp. Teor. Fiz.
\textbf{105}, 747 (1994) [Sov. Phys. JETP \textbf{78}, 401
(1994)].

\bibitem{SZB04}
K. V. Samokhin, E. S. Zijlstra, and S. K. Bose, Phys. Rev. B
\textbf{69}, 094514 (2004) [Erratum: \textbf{70}, 069902(E)
(2004)].

\bibitem{Book}
V. P. Mineev and K. V. Samokhin, \emph{Introduction to
Unconventional Superconductivity} (Gordon and Breach, London,
1999).

\bibitem{FAKS04}
P. A. Frigeri, D. F. Agterberg, A. Koga, and M. Sigrist, Phys.
Rev. Lett. \textbf{92}, 097001 (2004) [Erratum: \textbf{93},
099903(E) (2004)].

\bibitem{SM08}
K. V. Samokhin and V. P. Mineev, Phys. Rev. B \textbf{77}, 104520
(2008).

\bibitem{LL3}
L. D. Landau and E. M. Lifshitz, \emph{Quantum Mechanics}
(Butterworth-Heinemann, Oxford, 2002).

\bibitem{Lebed06}
A. G. Lebed, Phys. Rev. Lett. \textbf{96}, 037002 (2006).

\bibitem{NMP}
K. V. Samokhin and M. S. Mar'enko, Phys. Rev. Lett. \textbf{97},
197003 (2006).


\end{thebibliography}
\end{document}